\documentclass[9pt,twocolumn,twoside]{opticajnl}
\journal{opticajournal} 
\setboolean{shortarticle}{true}
\usepackage{lineno}
\usepackage{graphicx}
\usepackage{dcolumn}
\usepackage{bm}
\usepackage{caption}
\usepackage{indentfirst}
\usepackage{subcaption}
\usepackage{physics}
\usepackage[justification=justified]{caption}
\usepackage{subcaption}

\title{Light shift suppression in a CPT magnetometer using linear polarization and double frequency interrogation}

\author[1,2]{M.A. Maldonado}
\author[1,2]{Yang Li}
\author[3]{James A. McKelvy}
\author[3]{Andrey Matsko}
\author[4]{Irina Novikova}
\author[4]{Eugeniy E. Mikhailov}
\author[2]{John Kitching}
\author[2,*]{Ying-Ju Wang}

\affil[1]{University of Colorado, Boulder, CO 80301, USA.}
\affil[2]{National Institute of Standards and Technology, Boulder, CO 80305, USA.}
\affil[3]{Jet Propulsion Laboratory, California Institute of Technology, Pasadena, CA 91109, USA}
\affil[4]{Physics Department, College of William \& Mary, Williamsburg, VA 23187, USA}
\affil[*]{ying-ju.wang@nist.gov}

\begin{abstract}
We demonstrate a suppression of the light shift in a Coherent-Population-Trapping (CPT) atomic magnetometer by using linearly polarized light and a differential measurement between magnetic resonances. The radio frequency that creates the optical sidebands for CPT quickly switches between two magnetic sensitive transitions and the magnetic field is extrapolated from the difference of the center frequencies of the magnetic resonances. Light shifts and common drifts like collisional shifts can be suppressed through careful choice of measured resonances and we show the light shift reduction by more than a factor of 20 compared to excitation with circular polarized light. Various limitations to the method are discussed.
\end{abstract}

\setboolean{displaycopyright}{false} 

\begin{document}

\maketitle

Highly sensitive atomic magnetometers are useful in many areas of research, ranging from biomedical imaging \cite{Kim2014, Colombo2016} and material characterization \cite{ROMALIS2011,Bevington2019} to tests of fundamental symmetries \cite{Zhou2017, Zhang2023}. Due to their simplicity, high dynamic range, and low power consumption, atomic magnetometers based on Coherent Population Trapping (CPT) have previously been employed in space missions \cite{Cheng2018, Pollinger2018} where stability and accuracy are critical \cite{Amtmann2024, connerney2015, Fratter2015}. 

The energy levels in atomic systems shift in the presence of optical fields. Temporal variations of the intensity, frequency or polarization of the optical fields used to excite atoms can cause errors in atomic sensors which exploit transition frequencies between atomic energy levels to measure physical quantities of interest. Such "light shifts" often significantly contribute to the long term instability of a variety of atom-based instruments. For example, the light shifts of the Zeeman sub-levels add noise to the measurement of magnetic fields \cite{Mathur1968,Cohen-Tannoudji1972}. Such low-frequency noise of vapor-cell-based magnetometers has been characterized at sub-mHz level \cite{Mateosa2015}. Understanding, and ultimately suppressing, this shift is therefore important in improving the long-term frequency stability of these types of instruments. 

For continuous wave (cw) clocks based on coherent population trapping, methods have been proposed to reduce light shifts by appropriate choice of the optical frequency \cite{Zhu2009, Gerginov2018}; the frequency or amplitude of the modulation that is used to create coherent sidebands of the laser fields \cite{Zhu2000, McGuyer2009,Levi2000,Mikhailov10}; or polarization of the probe light \cite{Liu2022}. Unlike CPT clocks, which typically measure the center frequency of a single resonance of a magnetic-insensitive transition, magnetometers using linearly polarized light can interrogate multiple resonances simultaneously and provide simple extraction of vector field information, i.e., field orientation \cite{Yudin2010,Cox2011,Maldonado2024} because the amplitude ratios between resonances depend only on the optical polarization angle with respect to the magnetic field. CPT magnetometers using linearly-polarized light are also expected to suppress light shifts and other temperature-dependent drifts in vapor cells, such as the collision shift and quadratic Zeeman shift \cite{Yudin2010}. In this work, we demonstrate light shift suppression of a CPT magnetometer by using a linear-polarized laser beam and differential frequency detection with direct digital synthesis (DDS) to quickly switch the radio frequency for fast interrogation on multiple resonances.

In the effective operator formalism \cite{Happer1967, Mathur1968}, light  shifts in atomic systems can be expressed in terms of the electric field $\textbf{E}=\varepsilon_0 e^{-i(\omega t+\hat{k}\cdot\hat{r})}\hat{\zeta}+c.c.$ and the polarizability operator $\alpha^L$ (L: scalar, vector and tensor) \cite{Happer1967} as: 
\begin{equation}
    \label{eq:LSoperator}
    \begin{split}
    \delta E=&\delta E_0+h\delta AI\cdot J-\mu\cdot\delta H+\delta\varepsilon_2 \\=&- \frac{1}{8}\mid \varepsilon_0\mid^2\sum_{L}(\hat{\zeta}^*\cdot\alpha^L\cdot \hat{\zeta}+\hat{\zeta}\cdot\alpha^{L\dagger}\cdot \hat{\zeta}^*).
    \end{split}
\end{equation}
Here $\delta E_0$ is a common shift of all sublevels that has no impact on the transition frequencies and $h\delta A I\cdot J$ is the scalar light shift that shifts all Zeeman levels within a given hyperfine manifold equally. When there exists a circular polarization component, the transition frequency has the Zeeman (vector) light shift contribution of $\mu\cdot\delta H$, where $\delta H$ is an effective magnetic field proportional to the light intensity. The final term $\delta\varepsilon_2$ is the tensor light shift and is usually small compared to the scalar and vector light shift due to the smaller excited-state hyperfine splitting. 

In a CPT magnetometer, two light fields illuminate the atomic ensemble in a $\Lambda$ configuration, $\textbf{E}_{1,2}=\varepsilon_{1,2}e^{-i(\omega_{1,2}t+\hat{k}_{1,2}\cdot\hat{r})}\hat{\zeta}_{1,2}+c.c.$, with their frequency difference approximately matching the ground-state hyperfine splitting $\nu_{HF}$ (6.834 GHz). In the presence of a magnetic field $B$, the frequency of the transition between two $\Lambda$-coupled Zeeman states $\ket{F_1, m_{F_1}}$ and $\ket{F_2, m_{F_2}}$ is given to first order in $B$ by \cite{Knappe1999}:
\begin{equation}
   \label{eq:CPTSpectrum}
\nu_R(n,\Delta m) = \frac{\delta\varepsilon_{n,\Delta m}}{h} \approx(g_J-g_I)\frac{\mu_B B}{8h}n+g_I \frac{\mu_B B}{h} \Delta m
\end{equation}
where $g_J$ and $g_I$ are the electronic and nuclear g-factors respectively, $\mu_B$ is the Bohr magneton and where $n=m_{F_1}+m_{F_2}$ and $\Delta m = m_{F_2}-m_{F_1}$. The resulting spectrum is a series of doublets with the coarse structure determined by the electronic Zeeman splitting (the first term to the right of the approximation sign in Eq.~\ref{eq:CPTSpectrum}) and each doublet split by the much smaller second term due to the nuclear magnetic moment.
A measurement of the difference in frequency $\Delta \nu^B$ between two resonances ($n_1$, $\Delta m_1$) and ($n_2$, $\Delta m_2$), can then precisely determine the magnitude of the magnetic field as
\begin{equation}
\begin{split}
\Delta \nu^B = [(g_J-g_I)\frac{\mu_B}{8h}(n_1-n_2) + g_I \frac{\mu_B}{h}(\Delta m_1 - \Delta m_2)]B.
\end{split}
\end{equation}

In presence of the light, the two photon resonance condition is further modified by the interaction of the optical fields with the atoms and the measured frequency shift becomes  
\begin{equation}
    \label{eq:dif}
    \Delta\nu_{meas}=\Delta\nu^B + \Delta\nu^l (\omega_1, \omega_2). 
\end{equation}
The additional term produces a systematic error and limits the sensitivity with which the magnetic field strength can be determined in the presence of polarization, frequency or amplitude noise on the optical field. For a particular $n$, the effective shift of the corresponding CPT transition $\nu^l_n$ is given by the differential light shift of the two coupled levels $m_{F_{1,2}}$, and can be separated into its three contributions as in  Eq.~\ref{eq:LSoperator} \cite{Hu17}:
\begin{equation}
\label{eq:S}
\begin{split}
    \nu^S_n=-(\frac{\varepsilon_2}{2})^2/h\{[ q\alpha^S_2(\omega_1)+\alpha^S_2(\omega_2)]\\-[ q\alpha^S_1(\omega_1)+\alpha^S_1(\omega_2)]\},
\end{split}
\end{equation}

\begin{equation}
\label{eq:V}
\begin{split}
    \nu^V_n=-(\frac{\varepsilon_2}{2})^2(\hat{k}\cdot\hat{B})\mathcal{A}/h\{\frac{m_{F_2}}{4}[ q\alpha^V_2(\omega_1)+\alpha^V_2(\omega_2)]\\-\frac{m_{F_1}}{2}[q\alpha^V_1(\omega_1)+\alpha^V_1(\omega_2)]\},
\end{split}
\end{equation}

\begin{equation}
\label{eq:T}
\begin{split}
    \nu^T_n=-(\frac{\varepsilon_2}{2})^2(3\mid\hat{\zeta}\cdot\boldsymbol{\hat{B}}\mid^2-1)/h\{ \frac{3m_{F_2}^2-6}{12}[q\alpha^T_2(\omega_1)\\+\alpha^T_2(\omega_2)]-\frac{3m_{F_1}^2-2}{2}[ q\alpha^T_1(\omega_1)+\alpha^T_1(\omega_2)]\}
\end{split}
\end{equation}
where $\hat{B}$ is the unit vector along the quantization axis. $\alpha_F^{S,V,T}(\omega)$ are the scalar, vector, and tensor polarizabilities of the ground state $F$ induced by the optical field with frequency $\omega$. The field's amplitude $\varepsilon_1^2$ is written as a fraction of $\varepsilon_2^2$ ($\varepsilon_1^2=q\varepsilon_2^2$), where $q$ is fixed for a given RF modulation index. The symbol $\mathcal{A}$ represents its degree of circular polarization ($\mathcal{A}=\pm1$ for $\sigma^\pm$-polarized light and $\mathcal{A}=0$ for linearly-polarized light). Only the vector $\nu^V_n$ and tensor $\nu^T_n$ parts contribute to $\Delta\nu^l$ in Eq.~\ref{eq:dif} when taking the differential shift between any two CPT transitions. The scalar component $\nu^S_n$  is a common shift that doesn't depend on the specific $m_F$ values and therefore vanishes when taking the difference of any pair of CPT resonance frequencies. 

\begin{figure}[h]
\centering
\includegraphics[width=0.70\linewidth]{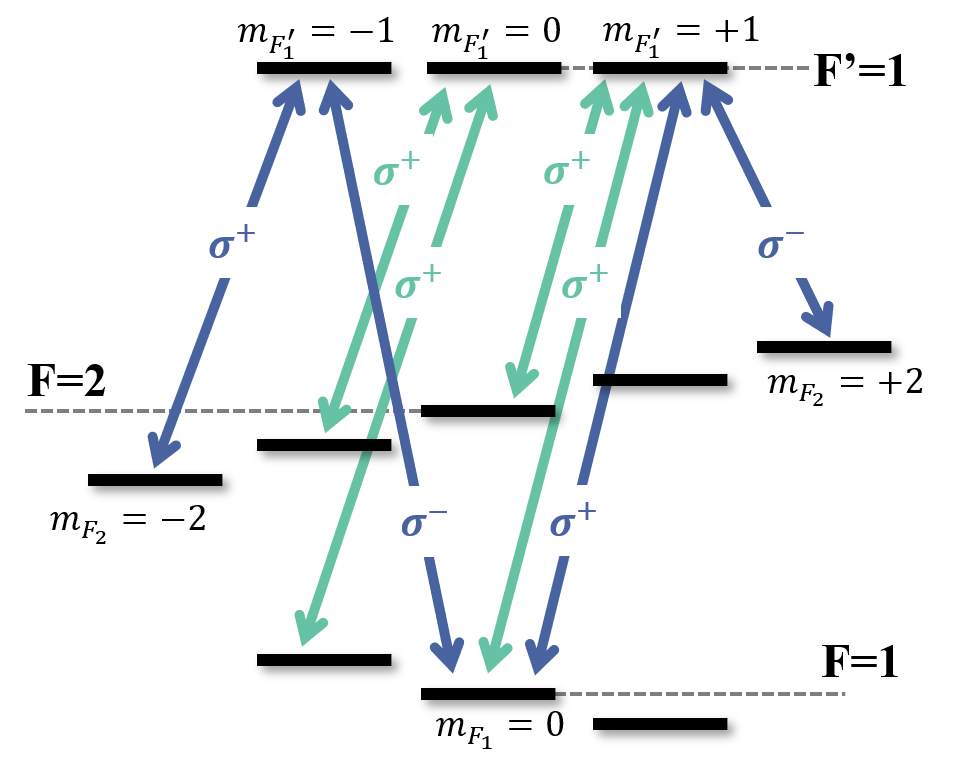}
\caption{\label{fig:level} Energy level diagram of the D1 line of $^{87}$Rb. The arrows represent the different $\Lambda$-schemes created with circularly (green) and linearly (blue) polarized light to excite the n=$\pm$ 2 transition.}
\end{figure}

For a CPT magnetometer with circularly-polarized light and atoms with nuclear spin 3/2, only two resonances are present in the spectrum and correspond to transitions $\ket{F=1,m_{F_1}=0} \leftrightarrow \ket{F=2, m_{F_2}=0}$ and $\ket{F=1,m_{F_1}=-1} \leftrightarrow \ket{F=2, m_{F_2}=-1}$, represented by green arrows in Fig.~\ref{fig:level}. Both the vector and tensor components of the light shift contribute here since $|\mathcal{A}|=1$ and each transition has different magnitudes of $m_{F_1}$ and $m_{F_2}$. When using linear polarization, $\mathcal{A}=0$ in Eq.~\ref{eq:V} and the vector light shift is completely suppressed for all CPT transitions. The tensor shift can be further canceled by measuring the differential shift of two transitions with $m_F$'s of the same magnitude. 
Such a pair exists on the D1 line of $^{87}$Rb, and are given by
$\ket{F=1,m_{F_1}=0}\leftrightarrow \ket{F=2,m_{F_2}=-2}$ (for $n=-2$, $\Delta m=-2$) and $\ket{F=1,m_{F_1}=0}\leftrightarrow\ket{F=2,m_{F_2}=+2}$ (for $n=+2$, $\Delta m=+2$) and are schematically represented by the blue arrows in Fig.~\ref{fig:level}. 

In this work, we show experimentally that the vector and tensor light shifts can indeed be suppressed by a factor of more than 20 using CPT excitation with linearly-polarized light and by taking the appropriate differences in resonance frequencies within the CPT spectrum.

\begin{figure}[h]
     \centering
     \includegraphics[width=0.45\textwidth]{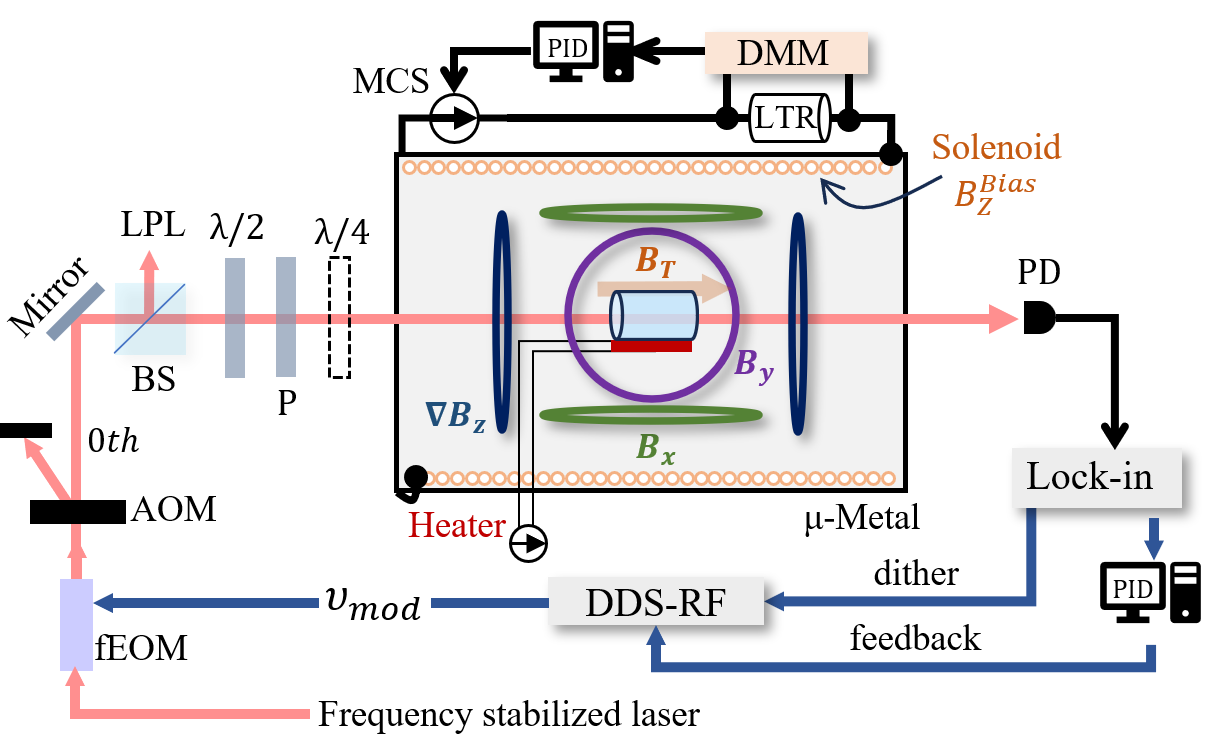}
     \caption{\label{fig:setup} Experimental Setup. LPL: laser power lock, PD: photo-detector, fEOM: fiber electro-optical modulator, BS: beam splitter, P: polarizer, LTR: low temperature coefficient resistor, MCS: main current source, DMM: digital multimeter.}
\end{figure}

To demonstrate the suppression of the vector light shift using linearly polarized light, we excite the CPT transitions at 795 nm in a heated glass-blown vapor cell with an inner volume of 1 cm$^3$, as shown in Fig. \ref{fig:setup}. A Ne buffer gas with 1.3 kPa of pressure is added to increase the coherence time of the atoms. The laser frequency is locked to the $F=2 - F'=1$ transition in a reference cell with an offset lock technique to compensate for the buffer gas shift of the excited transitions. The bi-chromatic field for the CPT process is generated using a fiber-coupled electro-optical modulator (fEOM). The two-photon resonance is satisfied using the carrier and 1st sideband at a modulation frequency of approximately $\nu_{mod}=6.8$ GHz $+\nu_{R}(n,\Delta m)+\nu^l$.  
The laser power is stabilized using the 0th-order output from an acousto-optic modulator (AOM).  The vapor cell is placed in a controlled magnetic field environment inside a multi-layer $\mu$-metal shield. The quantization axis in the $\hat{z}$ direction is defined by a longitudinal magnetic field \textbf{$B^T$}, produced by a solenoid, generating a longitudinal field $B_z^{Bias}$, and two sets of perpendicular saddle coils, generating transverse fields $B_x$ and $B_y$, with which the direction of the applied field can be fine-tuned to be along the wave-vector of the light, $\hat{k}$. An extra pair of coils in an anti-Helmholtz configuration generates a field gradient $\nabla B_z$ with which magnetic gradients internal to the shield can be compensated. The solenoid current (MCS) is monitored and stabilized by passing it through a low-temperature-coefficient resistor (LTR-Fig.\ref{fig:setup}) and measuring the resulting voltage using a high-performance digital multimeter (DMM-Keysight 3458A-Fig.\ref{fig:setup}). A polarizer before the cell suppresses the polarization components perpendicular to its optical axis by more than a factor of 40 dB. A photodetector (PD) after the cell measures the transmitted light. 

We determine the resonance frequency of the CPT transitions by slowly (230 Hz) dithering the frequency $\nu_{mod}$ of the RF field, and locking it to the zero-crossing of the dispersive error signal generated by a lock-in amplifier.  A direct digital synthesizer (DDS) is used to hop the modulation frequency $\nu_{mod}$ between two CPT resonances at a rate of $\approx$ 1 s to further determine the differential frequency shift $\Delta\nu_{meas}$ between the two transitions. This method helps to remove the effects of common drifts on the CPT spectrum during the measurement time.

When the light propagation direction is perfectly aligned with the magnetic field, a linear polarization excites both $\sigma^+$ and $\sigma^-$ transitions. At high magnetic fields, a doublet forms at each $n=\pm 2$ resonance, created by the $\Delta m=0$ (for $\sigma_+/\sigma_+$) and $\Delta m = 2$ (for $\sigma_+/\sigma_-$) transitions in Fig.~\ref{fig:level}. As described above, the doublet in the CPT spectrum is a result of the nuclear magnetic moment, which slightly modifies the linear Zeeman shift in each hyperfine manifold by a different amount \cite{Knappe1999}. At moderate magnetic fields, these resonances overlap and create ambiguity in the measurement of the resonance frequency. To resolve the $n=\pm2$ transitions, a strong magnetic field $B^T$ of 684 $\mu$T is applied, which splits the doublets by approximately 26 kHz and 12 kHz respectively. The field gradients at such a field strength do not significantly broaden the resonances, thereby maintaining reasonable resolution for the measurement of the resonance frequencies. 

\begin{figure}[h]
     \centering
     \includegraphics[width=0.40\textwidth]{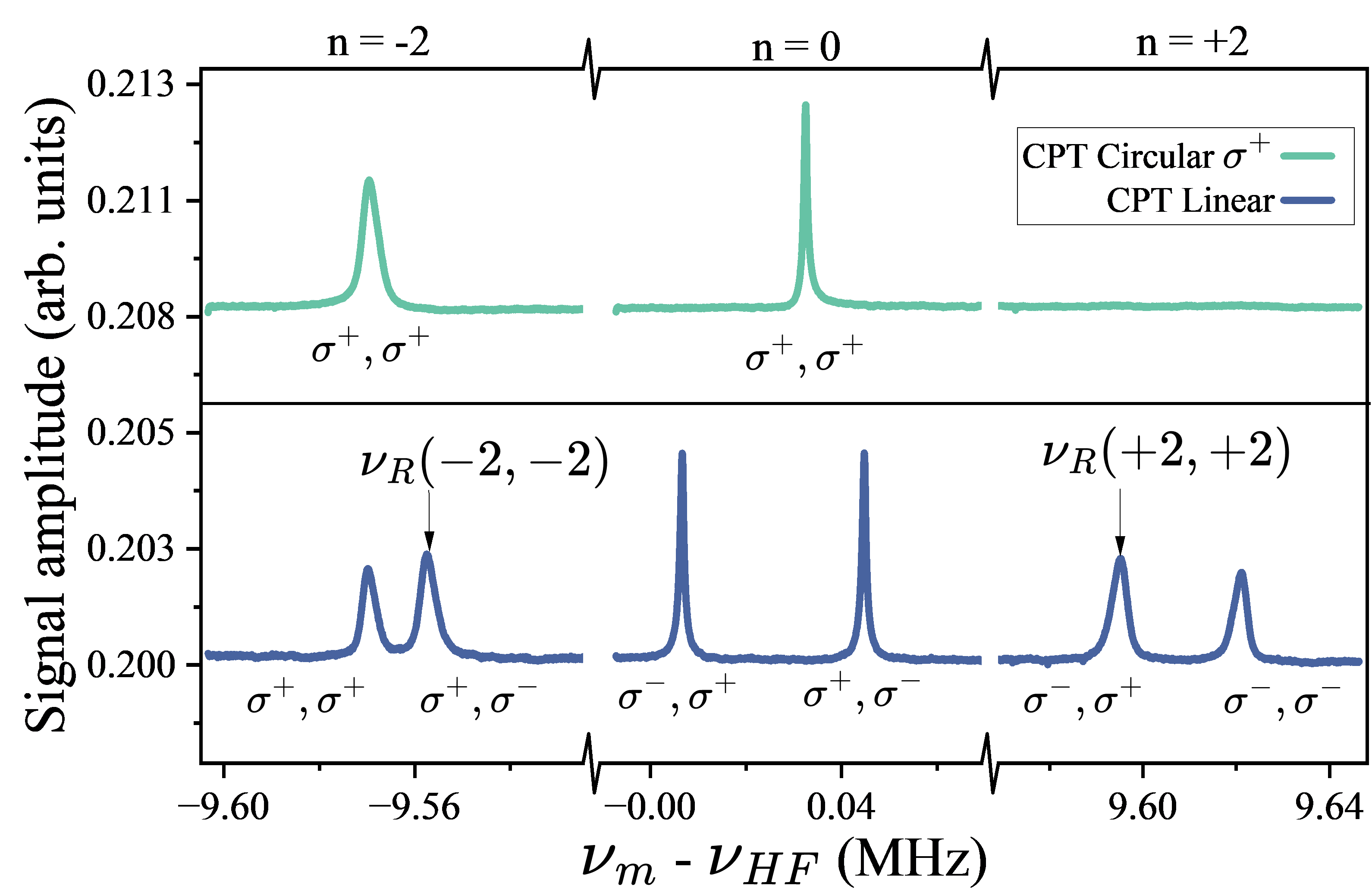}
     \caption{\label{fig:cpt} CPT transitions excited by circular and linear polarizations.}
\end{figure}

The CPT spectrum produced by linearly polarized light at this field is shown by the blue trace in the Fig.~\ref{fig:cpt}.   
There is an overall shift of all the resonances with respect to the hyperfine frequency of the unperturbed clock transition $\nu_{HF}$ due to the presence of the buffer gas. The linewidth of the n = $\pm$ 2 peaks is broadened to 3.7 times that of the clock n = 0 transitions (approximately 1 kHz) by the residual magnetic field gradient (roughly 200 nT/cm). Also shown in Fig.~\ref{fig:cpt} is the CPT spectrum obtained using circularly-polarized light (green trace). Here the $\Delta m = 2$ resonances are not excited, leaving only the singlet $\Delta m = 0$ resonances.

\begin{figure}[h]
     \centering
     \includegraphics[width=0.40\textwidth]{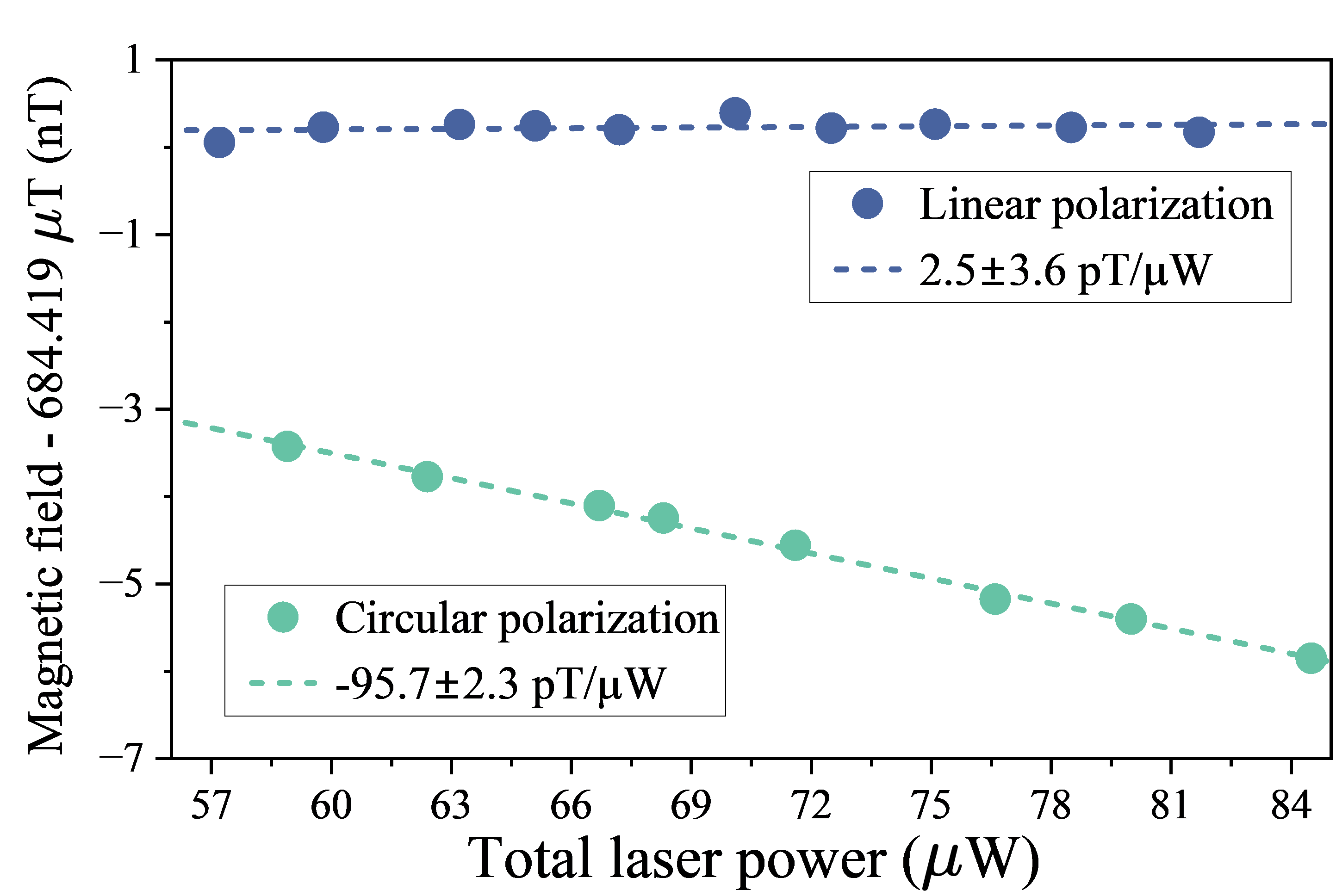}
     \caption{\label{fig:LSmeasurement} Differential light shift for circular (green) and linear polarization (blue).}   
\end{figure}

For circularly polarized light, we measure a differential shift between the  $n=0$ and $n=-2$ transitions that varies with optical power. The corresponding value of the magnetic field is calculated from the measured frequency shift using the Breit-Rabi equation and the result is plotted in green in Fig.~\ref{fig:LSmeasurement}. The scalar light shift is canceled in the differential frequency measurement, so the vector and tensor components are the major contributions to the measured shift, with a slope given by -95.7 $\pm$ 2.3 pT/$\mu$W. In the case of linear polarization, the differential shift is measured between the pair of the $n=\pm2$ transitions, marked as $\nu_R(\pm2,\pm2)$ in Fig.~\ref{fig:cpt} and represented with blue arrows in Fig. ~\ref{fig:level}. 

We observe no measurable change in extracted magnetic field over the entire range of laser powers, consistent with complete suppression of the vector light shift within the resolution of our measurements. A linear fit to the data returns a light shift coefficient of 2.51 $\pm $ 3.61 pT/$\mu$W, consistent with zero. The vector light shift vanishes with linear polarization, and the scalar as well as tensor light shift also cancel when measuring the difference in transition frequency of $n=\pm2$. The CPT magnetometer with linear polarization is therefore shown to be much less sensitive to changes in laser power.

The cancellation of the light shift is more than a factor of 20 when using linear polarization and a differential measurement between resonances, in comparison with the case of circular polarization, even at the highest laser powers. In addition, when we extrapolate the magnetic field to zero intensity in each polarization case, the resulting measured fields are equal within 2 nT, given by 684.4212$\pm$0.0002 $\mu$T for the circular polarization and 684.4191$\pm$0.0003 $\mu$T for the linear polarization.

The suppression of the light shift observed here for linear polarization implies that fluctuations in laser power should have a substantially reduced effect on the magnetometer output, thereby improving the long-term magnetometer stability and potentially its accuracy also. However, there are several factors that could limit this enhancement of performance. At higher laser powers and weak fields, power broadening results in overlap and distortion of the resonances as the doublets merge into a single peak, which contaminates the measurement of their individual resonance center frequencies. In addition, polarization noise that changes the optical pumping and hence the population ratio of states, would change the relative signal amplitude within the merged doublet and shift center frequency, thus contributing corresponding field instability. 

Stress-induced birefringence of optical components between the polarizer and the atoms, which adds a small degree of circular polarization to an otherwise linearly polarized beam, also shifts the resonance center frequency and affects the light-shift stability.   
\begin{figure}[h]
     \includegraphics[width=0.46\textwidth]{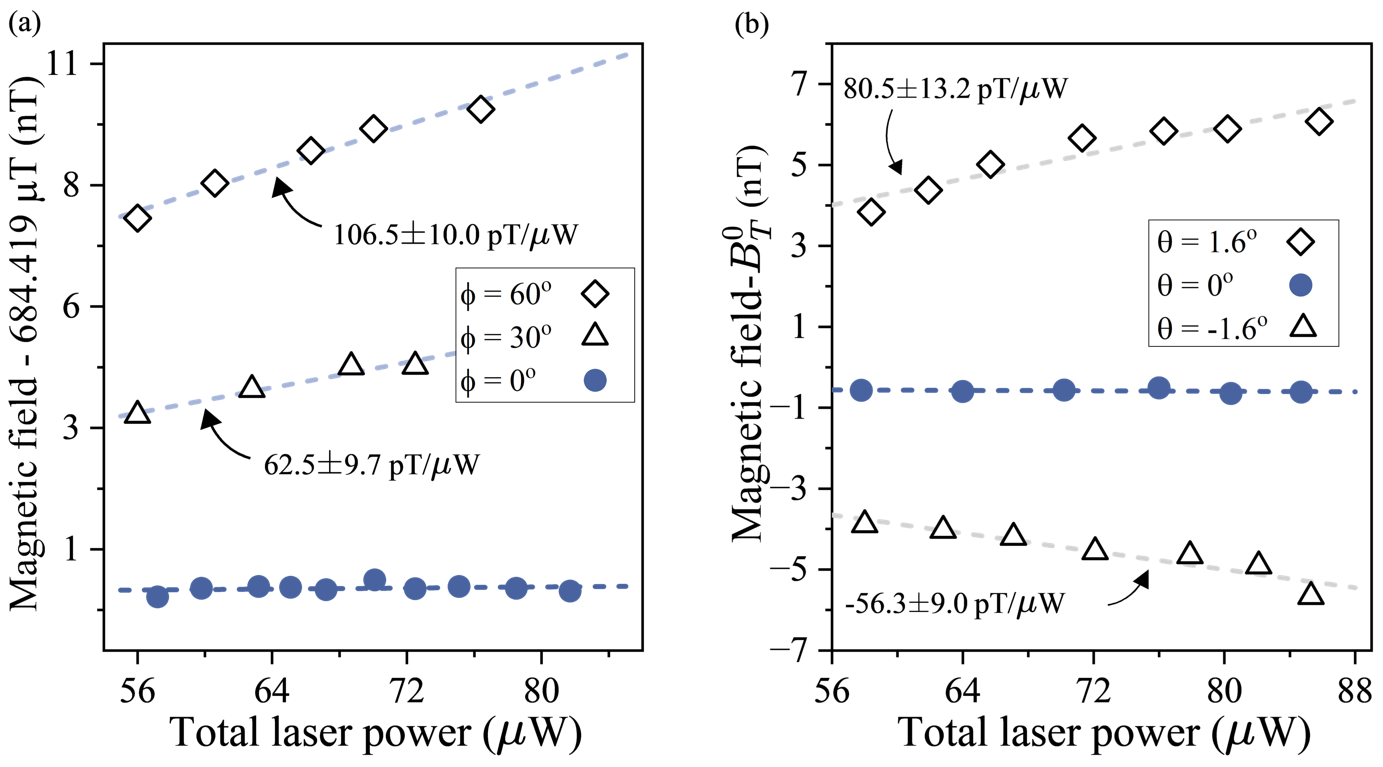}
     \caption{\label{fig:lightshiftdependence} (a) Light shift dependence on the optical axis alignment (b) Light shift dependence on the $\hat{k}$ and $\hat{B_T}$ alignment.}   
\end{figure}
When rotating the linear polarization axis with the laser beam propagating along $\hat{B}_T$, we observed that the slope of the light shift changed with the angle $\phi$ of linear polarization. This is due to a small birefringence of the input window of the vapor cell. The light shift coefficient varies by about 1.8 (pT/$\mu$W)/degree when the linear polarization is close to the optical axis of the cell window as shown in figure \ref{fig:lightshiftdependence}(a).

The light shift coefficient also changes noticeably by about 43 (pT/$\mu$W)/degree if there is a misalignment of the laser beam with respect to the magnetic field as shown in figure \ref{fig:lightshiftdependence}(b). Here we changed the angle $\theta$ between $\hat{k}$ and the bias field by changing the magnitude of the perpendicular field $\hat{B_x}$. The new bias field has a magnitude of $B_T^0=\sqrt{B_T^2+B_x^2}$ with a relative angle $\theta=tan^{-1}(B_x/B_T)$ with respect to its original direction $\hat{B_T}$. The change on the light shift coefficient is a result of an additional excitation and resonance contribution from the $\pi$ transition that would not exist if the alignment of laser beam with respect to the magnetic field were perfect ($\theta=0$). 

The measurement of the stability of our current magnetometer is largely limited by the noise in the bias magnetic field which is estimated to be about 200 pT/$\sqrt{Hz}$ at 10 mHz. Nevertheless, our work demonstrates that the light shift of CPT magnetometers can be suppressed by more than an order of magnitude with linear polarized light fields and differential measurement between $n=\pm2$ resonances with same $\Delta m$. This suppression of the light shift is expected to improve the low frequency noise of the magnetometer, which can be established with an improved bias field current source or gradiometric measurement. Such a stable magnetometer is expected to benefit applications in space science and electrical metrology where long-term magnetometer stability and accuracy are desired.

\begin{backmatter}
\bmsection{Funding}
The work is supported by Defense Advanced Research Projects Agency (DARPA) Science of Atomic Vapors for New Technologies (SAVANT) program. M. A. M, Y. L., J. K., and Y.-J. W. are also supported by National Institute of Standards and Technology (NIST). The research conducted by J. A. M. and A. M. was carried out at the Jet Propulsion Laboratory, California Institute of Technology, under a contract with the National Aeronautics and Space Administration (80NM0018D0004).
\bmsection{Acknowledgment} 
The authors thank Issac Fan for his experiment assistant and helpful comments.
\bmsection{Disclosures}
The authors declare no conflicts of interest. Mention of commercial products is for information only and does not imply recommendation or endorsement by NIST.
\bmsection{Data Availability Statement} 
Data underlying the results presented in this Letter are not publicly available at this time but may be obtained from the authors upon reasonable request.
\end{backmatter}

\bibliography{sample.bib}
\end{document}